\documentstyle[12pt]{article}
\begin{document}

\begin{titlepage}

\title{Tricritical Phenomena in a Z(3) Lattice Gauge Theory}

\author{N.S.~Ananikian\thanks{E-mail address: nanan@tpd.erphy.armenia.su}
\\[2mm]
{\small \sl Department of Theoretical Physics,
            Yerevan Physics Institute,} \\
{\small \sl Alikhanian Br. 2, 375036 Yerevan, Armenia} \\[2mm]
and \\[2mm]
R.R.~Shcherbakov\thanks{E-mail address: shcher@thsun1.jinr.dubna.su}
                \thanks{On leave of absence from
               Department of Theoretical Physics,
               Yerevan Physics Institute, Armenia} \\[2mm]
{\small \sl Bogoliubov Laboratory of Theoretical Physics,} \\
{\small \sl JINR, 141980 Dubna, Russia}
}

\date{\nonumber}
\maketitle

\begin{abstract}
The Z(3) gauge model with double plaquette representation
of the action on a generalized Bethe lattice of plaquettes
is constructed.
It is reduced to the spin-1 Blume-Emery-Griffiths (BEG) model.
An Ising-type critical line of a second-order phase transition
ending in the tricritical point is found.
\end{abstract}
\bigskip

\thispagestyle{empty}
\end{titlepage}

\newpage
\setcounter{page}1
\normalsize

A continuum limit of a lattice gauge theory may be
constructed in the points of a second-order phase transition,
since infinite range correlations allow us to wipe out lattice
effects.

At a tricritical or multicritical points the number of relevant
couplings is larger than at the vicinity of the second-order
critical line. Therefore the existence of such points in the
lattice gauge theories opens a possibility for further
non-trivial continuum limits.

The purpose of this paper is a search for multicritical points
in the strong-coupling region of the Z(3) lattice gauge theory.

According to Kogut's formulation of a gauge Potts model on a
lattice~\cite{Kogut}, we constructed the Z(3) gauge model
on the generalized Bethe lattice of plaquettes~\cite{Ananikian,Akheyan}
with double plaquette (window) representation of the action.
The choice of this mixed action
allowed us to connect it with the Hamiltonian of  the  spin-1
BEG model~\cite{BEG} and to receive the $\lambda$-line of the
second-order phase transition ending in the tricritical point.

This lattice is a generalization of the Cayley tree. For our purpose
we consider it only with coordination number (the number of plaquettes
coming out from one link) 2. The dual lattice is constructed
by joining each nearest centers of plaquettes and as a result we get
the usual Bethe lattice with coordination number 4.

Dual lattices for the generalized Bethe ones with coordination numbers
higher than 2 are more complicated hierarchical lattices. For
example, if we consider the generalized Bethe lattice with coordination
number 3 then the dual lattice will be a Husimi lattice~\cite{Husimi}.

The enlarged lattice gauge actions involving new double plaquette
interaction terms
were proposed and studied in $3d$ and $4d$ by
Edgar~\cite{Edgar}, Bhanot~et al~\cite{Bhanot}. The $2d$
version of one of these lattice gauge models with
Z(2) gauge symmetry formulated on the planar rectangular
windows was investigated by Turban~\cite{Turban}.
This model with pure gauge action had been reduced to the
usual spin-$\frac{1}{2}$ Ising model on the square lattice
and the point of a second-order phase transition was found.

The model is considered
on the generalized Bethe lattice in terms
of the bond variables $U_b$ which take their values in the Z(3),
the group of the third roots of unity.
Let $U_{p_i}=\prod_{b\in\partial p}U_b$ denote the product of
$U_b$'s around an elementary plaquette $i$.

The gauge action of the model is
\begin{equation}
\label{sg}
S_{Gauge}(\beta_{2g},\beta_{2g}',\beta_{g}) =
            S_{p\,p} + S_{p}\,,
\end{equation}
where
\begin{displaymath}
S_{p\,p} = -\sum_{<p_i\,p_j>}
\left\{\beta_{2g} \left( \delta_{U_{p_i},1}\delta_{U_{p_j},1}+
                         \delta_{U_{p_i},z}\delta_{U_{p_j},z} \right)+
 \beta_{2g}' \left( \delta_{U_{p_i},1}\delta_{U_{p_j},z}+
		         \delta_{U_{p_i},z}\delta_{U_{p_j},1} \right)
\right\}\,,
\end{displaymath}
\begin{displaymath}
S_{p} = \beta_{g}\sum_{p_i}
\left( \delta_{U_{p_i},1}+\delta_{U_{p_i},z}\right)\,,
\end{displaymath}
$U_{p_i}$ denotes the usual plaquette variable, the product of link
gauge fields $U_{x,\mu}$ around an elementary plaquette.
The first summation goes over all nearest-neighbor plaquettes and
the second one is over all plaquettes of the lattice,
$z=\exp(i\frac{2\pi}{3})\in{\rm Z(3)}$.

Introducing spin variables $S_i$ in the sites of the dual lattice
such that
\begin{equation}
\label{ss}
\begin{array}{c}
S_i  =\delta_{U_{p_i},1}-\delta_{U_{p_i},z}\,,\\
S_i^2=\delta_{U_{p_i},1}+\delta_{U_{p_i},z}
\end{array}
\end{equation}
the action~(\ref{sg}) becomes
\begin{equation}
\label{sspin}
S_{Spin}(\beta_{2g},\beta_{2g}',\beta_{g})=
-\sum_{<ij>}\left\{
\frac{\beta_{2g}-\beta_{2g}'}{2} S_i S_j +
\frac{\beta_{2g}+\beta_{2g}'}{2} S_i^2 S_j^2
\right\} +
\beta_g\sum_i S_i^2
\end{equation}
in which we recognize the Hamiltonian multiplied by $1/k_BT$
of the well known BEG model~\cite{BEG}.

The corresponding partition function of the model~(\ref{sg}) on the
generalized Bethe lattice is
\begin{equation}
\label{zgauge}
Z_{Gauge}(\beta_{2g},\beta_{2g}',\beta_{g})=
\sum_{\{U\}}\exp\left[
-S_{Gauge}(\beta_{2g},\beta_{2g}',\beta_{g})\right]
\end{equation}
where the sum is taken over all possible configurations of
the gauge variables $\{U\}$. This partition function can be rewritten
in terms of the spin variables $S_i$ defined in the sites of the dual
lattice
\begin{equation}
\label{zgt}
Z_{Gauge}= 3^{N}Z_{Spin}^{Dual}
\end{equation}
where
\begin{displaymath}
Z_{Spin}^{Dual}=
\sum_{\{S\}}\exp\left[
-S_{Spin} \right]
\end{displaymath}
A factor $3^{N}$ has been included in the equation~(\ref{zgt}) to
take into account the difference between the number of gauge $\{U\}$
and spin $\{S\}$ configurations, since for each spin configuration
with N sites we have $3^{N}$ identical gauge ones.

A gauge invariant quantity $<\delta_{U_{p_i,1}}+\delta_{U_{p_i,z}}>$
is the order parameter of the model.

The BEG model on the usual Bethe lattice have been exactly solved
in~\cite{Avakian,Shcher}, where the $\lambda$-line of the
second-order phase transition ending in the tricritical point
have been found.
Thus we can rewrite these solutions in terms of the gauge couplings.
For the $\lambda$-line we have
\begin{equation}
\label{critic}
\exp (\beta_g^{\lambda})= 2(b-u_0)(u_0+1)^3u_0^{-1}\,,
\end{equation}
\begin{equation}
\label{parameter}
<\delta_{U_{p_i,1}}+\delta_{U_{p_i,z}}>_{\lambda} =
\frac{u_0(1+u_0)}{b+u_0^2}
\end{equation}
where
$$b=\exp \left(\frac{\beta_{2g}+\beta_{2g}'}{2}\right)
\cosh \left(\frac{\beta_{2g}-\beta_{2g}'}{2}\right) -1\,,$$
$$a=b/\left(\exp \left(\frac{\beta_{2g}+\beta_{2g}'}{2}\right)
            \sinh \left(\frac{\beta_{2g}-\beta_{2g}'}{2}\right)\right)\,,$$
$$u_0=a/(3-a)$$

The tricritical point is determined by the following expression
\begin{equation}
\label{Tri}
\frac{u_0+1}{b-u_0}=2+\frac{1}{8u_0}\,.
\end{equation}

Fig.1 shows the phase diagram of the Z(3) gauge model for the value
$\beta_{2g}'/\beta_{2g}=\frac{1}{2}$. The $\lambda$-line (solid line)
of the second-order phase transition starts at
$(\beta_{g}=-\infty,\beta_{2g}=2\log 2)$, and finishes at the
tricritical  point A which exists for the cases
$-\infty<\beta_{2g}'/\beta_{2g}<0.5478$. The tricritical point A
cuts the $\lambda$-line and separates the second-order phase transition
from the first-order one.

At the end we want to write down the tricritical exponents:
\begin{equation}
\label{exp1}
<\delta_{U_{p_i,1}}+\delta_{U_{p_i,z}}>-
<\delta_{U_{p_i,1}}+\delta_{U_{p_i,z}}>_{\lambda} \sim
|\beta_g - \beta_g^{Tri}|^{\frac{1}{2}}\,,
\end{equation}
\begin{equation}
\label{exp2}
<\delta_{U_{p_i,1}}+\delta_{U_{p_i,z}}>-
<\delta_{U_{p_i,1}}+\delta_{U_{p_i,z}}>_{\lambda} \sim
|\beta_{2g} - \beta_{2g}^{Tri}|^{}\,.
\end{equation}

We constructed the Z(3) gauge lattice model with double
plaquette (window) representation of the action and showed
that this model is dual to the spin-1 BEG one. Using the
exact solution of the BEG model on the Bethe lattice
we found the line of the second-order phase transition ending
in the tricritical point.

\bigskip

We would like to thank R.~Flume, K.~Oganessyan
and A.~Akheyan for fruitful discussions.

One of us (R.R.S.) wishes to thank V.B.~Priezzhev for the hospitality
extended to him at the Bogoliubov Laboratory of Theoretical Physics
of Joint Institute for Nuclear Research where this work has been done.

\medskip
This work was partially supported by German Bundusministerium f\"ur
Furschung und Technologie under the grant No. 211-5291 YPI and
by the grant INTAS-93-633.


\newpage

\ \ \

\vspace{6in}
\begin{center}
Figure 1.
\end{center}
\medskip
\noindent
{\small
Phase diagram of the Z(3) lattice gauge theory.
The $\lambda$-line of the second-order phase transition
(solid line) and the line of the first-order phase transition
(dashed line) meet together at the tricritical point A.
These lines separate the plane of the coupling constants
$(\beta_g,\beta_{2g})$ onto two regions
(ordered and disordered).

\end{document}